\documentclass[preprint,amsmath,amssymb]{revtex4}
\usepackage{graphicx}
\usepackage{tabularx}
\usepackage{makecell}
\usepackage{dcolumn}
\usepackage{adjustbox}
\usepackage{hyperref}
\usepackage{xcolor}
\usepackage{bm}

\begin{document}

\title{Electrodynamics in curved spacetime : Gravitationally-induced constitutive equations and the spacetime index of refraction}
\author{${\rm F. \;Masghatian}$ \footnote{Electronic
address:~fa.masghatian@gmail.com} ${\rm M. \;Esfandiar}$\footnote{Electronic
address:~esfandiar830@gmail.com}  and ${\rm M.\;Nouri-Zonoz}$ \footnote {Electronic address:~nouri@ut.ac.ir, (corresponding author)} }
\affiliation{Department of Physics, University of Tehran, North Karegar Ave., Tehran 14395-547, Iran.}

\begin{abstract}
Previous studies on spacetime index of refraction are mostly restricted to the static spacetimes. In this study we fill this gap by introducing the refractive index for stationary spacetimes, employing three different approaches. These include applying Fermat's principle, employing the classical definition of refractive index, and using gravitationally-induced constitutive equations. These calculations are carried out in both threading and slicing spacetime decomposition formalisms, and their equivalence is established. We discuss possible applications of the spacetime index of refraction, especially in the study of light trajectories and their characteristics in stationary spacetimes. More specifically we show that there is a gravitational analog of a toroidal moment, and discuss a possible gravitational analog of nonreciprocal refraction in materials with a toroidal domain wall.
\end{abstract}
\maketitle
%%%%%%%%%%%%%%%%%%%%%%%%%%%%%%%%%%%%%%%%%%%%%%%%%%%%%%%%%%%%%%
\section{Introduction}
One of the first studies of electromagnetic fields and Maxwell equations in curved spacetimes is due to Landau and Lifshitz in their classical treatise ``{\it classical theory of fields}'' \cite{LL}. In their brief study, by a suitable definition of electromagnetic fields in a {\it general} spacetime, they find gravitationally-induced constitutive equations and Maxwell equations in 3-vector language which are similar in form to their flat spacetime counterparts in curvilinear coordinates. Also, in the same reference, they derive Fermat's principle for rays propagating in a {\it stationary} spacetime in the context of threading formulation of spacetime decomposition, but do not go further to define the spacetime index of refraction, not even for the static spacetime. Misner and Wheeler, employing the slicing decomposition formalism and exterior calculus wrote down the vacuum Maxwell equations \cite{MW}. This was followed by the ADM study of Maxwell equations in the same context but in the presence of charges \cite{ADM}. \\
Around the same time Plebanski studied electromagnetic waves in gravitational fields {\cite{Pleb}, and in particular wrote Maxwell's equations in a decomposed 3-vector form and studied their analogy to the formulation of electrodynamics in  macroscopic media in flat spacetime. He also introduced almost the same gravitationally-induced  constitutive equations as in \cite{LL}.
In the early 1970's, first de Felice \cite{Defel}, and later Ellis \cite{Ellis} gave a covariant formulation of the electromagnetic fields projected along the 4-velocity of an observer. De Felice used Plebanski's formulation to study gravitational fields as media assigned with an index of refraction. Indeed their formalism based on the decomposition of the electromagnetic field tensor relative to a given observer's 4-velocity, generalizes the field decomposition in  threading  and slicing formalisms as two special cases corresponding to the fundamental (or Lagrange), and fiducial (or Eulerian) observers respectively \cite{Velocity}. \\
The analogy between a spacetime, and a medium with respect to the light propagation in geometric optics approximation has a long history going back to the studies of Gordon \cite{Gordon}, and Tamm \cite{Tamm} in the 1920's. This reciprocal opto-geometric relation between optical characteristics of materials, and the behavior of null rays in spacetime geometry has been recently developed as a basis for the so called {\it transformation optics} \cite{Pendry, Leon}. \\
Here we will introduce a systematic derivation of three-dimensional (3D) Maxwell equations in a stationary spacetime, employing both the threading, and slicing decomposition formalisms. Also we find gravitationally-induced constitutive equations in both formalisms, and use them to introduce the spacetime index of refraction. We show that spacetime refractive index, being dependent on the direction of propagation, allows for the gravitational analogs of nonreciprocal refraction, and light paths in the so called toroidal matter. Our study is based on geometric optics approximation, which is the working assumption in  many studies considering light propagation in curved spacetimes. In flat spacetime, the geometric optics corresponds to the approximation in which the wavelength is negligiblly small, i.e $\lambda \rightarrow 0$. More specifically in this approximation the wavelength is much smaller than the scale over which the amplitude changes. This scale could be taken as the wavefront's radius of curvature. In other words in this limit the waves can be locally treated as plane waves. In curved spacetime there is a second scale, the spacetime radius of curvature, which needs to be taken into account if treating light as rays is going to work. So in curved spacetime the geometric optics assumption corresponds to wave propagation with wavelength much smaller than these two scales \cite{MTW}. In other words this approximation is consistent with considering light rays as null geodesics of the underlying
spacetime.\\
The outline of the paper is as follows: In section II we briefly introduce both the threading and slicing spacetime decomposition formalisms.
In section III we show how one can assign an index of refraction to the stationary spacetimes, using two different approaches, in the threading decomposition formalism. In section IV, using the same formalism, we find Maxwell equations, and the gravitationally-induced constitutive equations in {\it stationary} spacetimes. In section V we use the constitutive equations to obtain the above-mentioned index of refraction, confirming our results in section III. In sections VI-VII we do the same calculations in slicing decomposition formalism, and in section VIII we stablish the equivalence between the results of the two formalisms by comparing their expressions for the spacetime refractive index. \\
In section IX, we discuss the possible applications of the spacetime refractive index in studying light ray trajectories in stationary spacetimes, and their characteristics. Specifically, we discuss ray trajectories in weak field stationary spacetimes, and show there is a gravitational analog of {\it nonreciprocal} refraction, and light  propagation in materials with toroidal moment.
We conclude with a summary, and discussion in the last section\\
{\bf Notations} : Our convention for indices is such that the
 Greek indices run from 0 to 3 while the Latin ones run from 1 to 3.
%%%%%%%%%%%%%%%%%%%%%%%%%%%%%%%%%%%%%%%%%%
\section{A brief introduction to spacetime decomposition formalisms}
The two main spacetime decomposition formalisms could be treated as the special cases of the decomposition along the 4-velocity of an observer. In this so called projection formalism, the metric of the spacetime which gives the 4D covariant distance between two nearby spacetime points, is locally projected along the observers 4-velocity and orthogonal to it. This is achieved by using the projection tensor $g_{\alpha\beta} = -\gamma_{\alpha\beta} + u_\alpha u_\beta$, so that the spacetime line element can be written in the following decomposed form
\begin{equation}\label{le}
ds^2 = d\tau^2 - dl^2 = (u_\mu dx^\mu)^2 - \gamma_{\mu\nu}dx^\mu dx^\nu.
\end{equation}
In what follow we will apply the above prescription to the 4-velocities of fundamental, and fiducial observers corresponding to the threading (or $1+3$) and slicing (or $3+1$) decomposition formalisms. Since in later sections we will explicitly mention which decomposition formalism is used, for the ease of writing, we employ the same notation for the 3D metric ($\gamma_{ij}$) in both formalisms. On the other hand the line element, and proper time are distinguished  by using indices T and S which stand for threading and slicing
 respectively.\\
{\bf 1) Fundamental observers} are assigned with the 4-velocity
\begin{equation}\label{vel1}
u^\mu= \frac{1}{\sqrt{g_{00}}} (1,0,0,0)\;\;\;\; ; \;\;\; u_\mu= \sqrt{g_{00}}(1, \frac{g_{0i}}{g_{00}})
\end{equation}
so that the line  element \eqref{le} projected along this 4-velocity reduces to
\begin{align}\label{metric1}
ds^2= {d\tau}_T^2- dl_T^2 = g_{00}(dx^{0}-g_{i}dx^{i})^{2} -\gamma_{ij}dx^{i}dx^{j}
\end{align}
in which $g_{i} = -\frac{g_{0i}}{g_{00}}$ is the so called {\it gravitomagnetic} potential, and
\begin{gather}
{d\tau}_T^2=g_{00}(dx^{0}-g_{i}dx^{i})^{2}\label{dt} \\
dl_T^2 = \gamma_{ij}dx^{i}dx^{j} =\left( -g_{ij} + g_{00} g_{i}g_{j} \right)dx^{i}dx^{j}.\label{dl}
\end{gather}
This form of the line element can also be obtained by using the radar signals sent between two nearby fundamental observers \cite{LL,LBNZ}. The above decomposition is valid in a general spacetime where both ${d\tau}$ and ${dl}$ could be time-dependent. On the other hand in a stationary spacetime one could show that \eqref{dt} and \eqref{dl}
are the so called  {\it synchronized} proper time, and the spatial line element (with 3D metric $\gamma_{ij}$) of the 3D {\it quotient space} ($\Sigma_3$) respectively \footnote{This 3D space is not necessarily a submanifold of the original 4D spacetime manifold  \cite{Geroch, Exact}.}. Indeed it was shown by Geroch that in the case of stationary spacetimes one can obtain the same form of spacetime decomposition by employing the  projection along the spacetime's {\it timelike Killing vector field}, namely $\xi^\alpha = \delta^\alpha_0$ \cite{Geroch}. In this case the fundamental observers are the so called Killing observers because their 4-velocity at each point is along the timelike Killing vector field.\\
Also for later use one can show that the following relations are held
\begin{equation}\label{gg}
\gamma^{ij} = - g^{ij} \;\;\; ; \;\;\;
-g = g_{00} \gamma  \;\;\; ; \;\;\;
g^i = \gamma^{ij} g_j = - g^{0i},
\end{equation}
in which $g$ and $\gamma$ are the determinants of $g_{\mu\nu}$ and $\gamma_{ij}$ respectively. \\
{\bf 2) Fiducial observers}, on the other hand, are assigned with the 4-velocity
\begin{equation}\label{vel2}
u_\mu= \frac{1}{\sqrt{g^{00}}} (1,0,0,0)\;\;\;\; ; \;\;\; u^\mu= \sqrt{g^{00}}(1, \frac{g^{0i}}{g^{00}}),
\end{equation}
which when substituted in \eqref{le} leads to the following familiar form of the line element in the well known slicing decomposition formalism
\begin{equation}\label{metric2}
ds^2={d\tau}_S^2- dl_S^2 = N^2 (dx^0)^2 -  \gamma_{ij} (N^i dx^0 + dx^i)(N^j dx^0 + dx^j),
\end{equation}
in which $N = \frac{1}{\sqrt{g^{00}}}$, and  ${N^i}= -{N^2} g^{0i}$ are the so called lapse function, and shift vector respectively. The proper time, and the spatial distance are now  defined as
\begin{gather}
{d\tau}_S^2=N^2 (dx^0)^2\label{dtf} \\
dl_S^2 = \gamma_{ij} (N^i dx^0 + dx^i)(N^j dx^0 + dx^j) \label{dlf},
\end{gather}
with
\begin{equation}\label{extras}
\gamma_{ij}= -g_{ij}\;\; ; \;\; -g = N^2~\gamma,
\end{equation}
where $g$ and $\gamma$ are the determinants of $g_{\mu\nu}$ and $\gamma_{ij}$ respectively. We note also the following relations
\begin{gather}
g_{00} = N^2 - N_i N^i \;\; ; \;\; N_i = -g_{0i}  \;\; ; \;\; \gamma^{ij} = - g^{ij} + \frac{1}{N^2} N^i N^j \label{efs}
\end{gather}
Using the above relations we can write the spacetime line element in terms of the contravariant components of the  3-metric in the following form
\begin{equation}\label{metric21}
ds^2={d\tau}_S^2- dl_S^2 = \frac{1}{N^2} (dx_0 -N^i dx_i)^2  -  \gamma^{ij} dx_i dx_j
\end{equation}
with
\begin{gather}
{d\tau}_S^2 = \frac{1}{N^2} (dx_0 -N^i dx_i)^2 \label{efss} \\
dl_S^2 = \gamma^{ij} dx_i dx_j \label{efs1}
\end{gather}
In what follows we are primarily interested in stationary spacetimes. This is so  because only in these spacetimes the finite distance obtained from integrating the spatial line elements \eqref{dl}, and \eqref{dlf} along a spatial curve has a time-independent meaning.
%%%%%%%%%%%%%%%%%%%%%%%%%%%%%%%%%%%%%%%%%%%%%%%%%%
\section{Index of refraction for stationary spacetimes}
One of the main concepts introduced in the study of analogies between a gravitational field and a medium, mentioned in the introduction, is the spacetime index of refraction, which is most studied in the case of static spacetimes.
Assigning an index of refraction to static spherically symmetric spacetimes in isotropic coordinates goes back to Eddington in his classic treatise ``{\it space, time, and gravitation}'' as one of the first books on GR \cite{Edding}. As pointed out in the introduction one can employ the mentioned analogies to design a (meta)material in flat space with the same index of refraction to reproduce  light paths mimicking those in the corresponding gravitational field.
The same procedure does not work for a given stationary spacetime as one can not transform its metric into isotropic coordinates, unless you restrict the analogy to a specific submanifold of the spacetime.  \\
But nonetheless, in stationary spacetimes one can assign an index of refraction to the spacetime by employing the {\it threading} decomposition formalism, and studying the behaviour of light rays in the corresponding spatial sector defined by the 3D metric and line element \eqref{dl} on $\Sigma_3$. In what follows we introduce the same definition for the index of refraction of a stationary spacetime, using two different methods.
%%%%%%%%%%%%%%%%%%%%%%%%%%%%%%%%
\subsection{Spacetime index of refraction from Fermat's principle}
Employing Fermat's principle in the context of {\it threading} spacetime decomposition formalism, we obtain a definition for the spacetime index of refraction for stationary spacetimes.
We start with the following form of Fermat's principle
\begin{equation}\label{FP}
\delta \int k_i dx^i = 0,
\end{equation}
where $k^\mu = \frac{dx^\mu}{d\lambda}$ is the wave 4-vector defined with respect to the affine parameter $\lambda$ along the light path. Using the  definition of wave vector, we can write the above relation in the following form
\begin{equation}\label{FP-1}
\delta \int k_i \frac{dx^i}{d\lambda} (\frac{d\lambda}{dl}) dl = \delta \int k_i k^i (\frac{d\lambda}{dl}) dl = 0,
\end{equation}
which upon  using the fact $k_\mu k^\mu = k_0 k^0 + k_i k^i = 0$, reduces to
\begin{equation}\label{FP-2}
\delta \int k^0 (\frac{d\lambda}{dl}) dl = 0,
\end{equation}
in which we used the fact that $k_0 = \frac{\omega_0}{c}$ is constant along the ray. Note that up to this point $d l$ could be the spatial line element in any decomposition formalism. To find $\frac{d\lambda}{dl}$ in threading formalism,  we start from  setting $ds^2 =0$ for null rays in \eqref{metric1}, which leads to
\begin{equation}
dl_T = \sqrt{g_{00}}(dx^{0}-g_{i}dx^{i}),
\end{equation}
then upon division by ${d\lambda}$, and using the definition of $k^\mu$ we obtain
\begin{equation}\label{dll}
\frac{dl_T}{d\lambda} = \sqrt{g_{00}}(k^0-g_{i}k^i) = \frac{k_0}{\sqrt{g_{00}}}.
\end{equation}
where in the last equation we used the fact that
\begin{gather}
k_0 = g_{0\mu} k^\mu = g_{00}(k^{0}-g_{i}k^{i}) \label{dk11}.
\end{gather}
Substituting \eqref{dll} back in \eqref{FP-2}, and replacing for $k^0$ from \eqref{dk11} we find,
\begin{equation}\label{FP-3}
\delta \int  {\sqrt{g_{00}}} {k^0} dl_T = \delta \int (\frac{1}{\sqrt{g_{00}}} +g_i \frac{\sqrt{g_{00}}}{k_0} k^i ) dl_T = 0.
\end{equation}
On the other hand substituting \eqref{dk11} in the relation $k_\mu k^\mu = 0$ we have
\begin{equation}\label{dk}
h(k^{0}-g_{i}k^{i})^{2} -\gamma_{ij}k^{i}k^{j}= \frac{{k_0}^2} {g_{00}} -\gamma_{ij}k^{i}k^{j} = 0,
\end{equation}
which shows, along with the relation $k^i = \frac{dx^i}{dl_T} (\frac{dl_T}{d\lambda})$, that we have
\begin{equation}\label{dk2}
k^i = \frac{{k_0}}{\sqrt{g_{00}}} \frac{dx^i}{dl_T} \equiv |{\bf k}| {{\hat k}^i}\;\; ; \;\; |{\bf k}| = \frac{dl_T}{d\lambda} = \frac{k_0}{{\sqrt{g_{00}}}}. \,
\end{equation}
Substituting the above relation in \eqref{FP-3}, we have the final result
\begin{equation}\label{FP-4}
\delta \int (\frac{1}{\sqrt{g_{00}}} + {\bf {g}} . {\bf{\hat k}})~dl_T \equiv \delta \int  n_T~d l_T = 0.
\end{equation}
which defines the spacetime index of refraction as
\begin{equation}\label{IRS2}
n_T = \frac{1}{\sqrt{g_{00}}} + {\bf {g}} . {\bf{\hat k}},
\end{equation}
with ${\bf \hat k}$  the unit wave vector along the ray direction. Obviously it reduces to $n = \frac{1}{\sqrt{g_{00}}}$, the refractive index  for static spacetimes.\\
An {\it important} point in passing is that this index of refraction is not defined with respect to spatial distance element in flat space, but in $\Sigma_3$ space on which, the  {\it spatial line element} $dl_T$ is defined by the corresponding 3D metric $\gamma_{ij}$.\\
To assign an index of refraction to a spacetime so that one can replace the effect of the gravitational field on light rays by a material medium in a flat spacetime, one has to find the relation between $dl$, and the spatial line element in flat spacetime. This could only be achieved by going to the isotropic coordinates, either in static spherically symmetric spacetimes \cite{NPF}, or in special hypersurfaces of a  stationary spacetime \cite{PFRN}. At this point, there are two remarks in order: \\
{\bf REMARK 1}: A by-product of the above study is the intoduction of  $d\lambda = \frac{c\sqrt{g_{00}}}{\omega_0} dl_T$ in Eq. \eqref{dll}. This gives  a general relation for the {\it affine parameter of null geodesics} in any stationary spacetime before explicitly specifying its null geodesics. For a recent discussion on the affine parameter of null geodesics refer to \cite{Visser}. Indeed, this relation between the affine parameter of a null geodesic, and the spatial line element in threading decomposition provided the  natural definition \eqref{dk2} for the  3D wave vector along a light ray.\\
{\bf REMARK 2}: Using the relations $k_i = g_{i\mu} k^\mu$, and  \eqref{dll}, along with the definition of the spatial metric $\gamma_{ij}$ in \eqref{dl}, one arrives at the following relation,
\begin{equation}\label{kd}
k_i= -\gamma_{ij} k^{j} - k_0 g_i.
\end{equation}
This relation  embodies a subtle point which needs to be stressed, and that is the important fact that $k_i$  {\it is not} a 3D covariant vector in $\Sigma_3$. So we should distinguish between spatial components of the covariant 4-vector $k_\mu$, namely $k_i$, and $\gamma_{ij} k^{j}$ which is the covariant counterpart of $k^i$ in $\Sigma_3$. We will employ this fact in later sections.
%%%%%%%%%%%%%%%%%%%%%%
\subsection{Spacetime refractive index from classical definition}
Here we show that there is an alternative way to derive the spacetime refractive index by employing its classical definition, namely as the ratio of light speed in vacuum to the light speed in a medium. Obviously here  the role of the medium is played by the gravitational field. To do so we start from the simple fact that for light rays the line element along its path vanishes, i.e
\begin{equation}\label{zero}
ds^2 =  g_{\mu\nu}dx^\mu dx^\nu = {d\tau}^2- dl^2 =  0
\end{equation}
from which we have
\begin{equation}\label{zero1}
dl = {d\tau} = {\tilde c} dt \;\; \Longrightarrow \;\; {\tilde c} = \frac{dl}{dt}
\end{equation}
where we have introduced the decomposition-based {\it coordinate speed of light} ${\tilde c}$.  This is the speed with which the light  traverses the spatial line element $dl$ (introduced in the corresponding decomposition formalism), in coordinate time $dt$. Employing the classical definition of index of refraction we have
\begin{equation}\label{Index}
n = \frac{c}{\tilde c} = \frac{c dt}{dl}=\frac{dx^0}{dl}.
\end{equation}
This is a general result, so in the case of threading formalism we replace the corresponding proper time and line element, $d\tau_T$, and $d l_T$ in \eqref{zero} to obtain
\begin{equation}\label{Index1}
d l_T = \sqrt{g_{00}}(dx^0 - g_i dx^i)\;\; \Longrightarrow \;\; 1= \sqrt{g_{00}}(\frac{dx^0}{dl_T} - g_i \frac{dx^i}{dl_T}).
\end{equation}
Now by using \eqref{Index}, and \eqref{dk2} in the above equation we find
\begin{equation}\label{Index2}
n_T = \frac{1}{\sqrt{g_{00}}} + {\bf {g}} . {\bf{\hat k}},
\end{equation}
which is the same relation obtained in \eqref{IRS2}.
%%%%%%%%%%%%%%%%%%%%%%%%%%%%%%%%%%%%%%%%%%%%%%%%%
\section{Electrodynamics in curved spacetimes and gravitationally-induced constitutive equations}
The covariant formulation of Electrodynamics in curved spacetime could be summarized in the following two Maxwell equations in terms of the electromagnetic field tensor
\begin{gather}
\partial_\beta F_{\mu\nu} + \partial_\nu F_{\beta\mu} + \partial_\mu F_{\nu\beta} = 0 \label{CME1} \\
\frac{1}{\sqrt{-g}}\partial_\nu (\sqrt{-g} F^{\mu\nu})= - \frac{4\pi}{c}j^\mu \label{CME2}.
\end{gather}
in which the components of the  {\it current} 4-vector $j^\mu$ should be defined with respect to the spatial volume element introduced in the corresponding spacetime decomposition formalism.\\
To develop further the idea of a ``{\it gravitational field as a medium}'' in the context of electromagnetic wave propagation in  curved backgrounds, one should be able to express the above equations in a 3D language, using a decomposition formalism. To this end one needs to know what are the counterparts of the electromagnetic fields $({\bf E},{\bf D},{\bf B},{\bf H})$ in a curved spacetime. To do so,
people either simply define all these fields in terms of the electromagnetic field tensor, apparently in an ad hoc way \cite{Pleb,LL}, or try to define at least  some of them (usually ${\bf E}\; {\rm and} \; {\bf H}$) through the projection of  the electromagnetic field tensor along a given 4-vector which could be the  4-velocity of an observer \cite{Defel, Ellis}. That is why these definitions normally differ from one author to another. This could be done in principle using either of the decomposition formalisms, though it is mostly carried out in the threading decomposition formalism. Finally by substituting these fields in the Maxwell equations \eqref{CME1}-\eqref{CME2} one obtains their desired 3D form. \\
Here we reverse the logic in the sense that we decompose the covariant Maxwell equations \eqref{CME1}-\eqref{CME2} by looking at their spatial, and spatio-temporal components in a given decomposition formalism and compare them with their flat spacetime counterparts in {\it curvilinear coordinates} to decide what should be taken as the definitions of the set of fields $({\bf E},{\bf D},{\bf B},{\bf H})$ in the corresponding curved spacetime. In this way there is no need to define them in terms of the projection of the field tensor along a given 4-vector which could be the  4-velocity of an observer, or a Killing vector field of the corresponding spacetime.\\
So for a given general spacetime (not necessarily stationary), first we look for the 3D forms of following spatial and space-time components of the source-free Maxwell equation \eqref{CME1},
\begin{gather}
\partial_b F_{mn} + \partial_n F_{bm} + \partial_m F_{nb} = 0 \label{CME11} \\
\partial_0 F_{mn} + \partial_n F_{0m} + \partial_m F_{n0} = 0 \label{CME12}
\end{gather}
and try to express them in 3D vector notation by applying the threading decomposition formalism. The tools we need in this regard include the Levi-Civita pseudotensor $\eta^{ijk}$, and the definitions of the curl, and divergence of 3-vectors in curved 3D space $\Sigma_3$, which  are given in the appendix A.\\
Having these tools, we start with equation \eqref{CME12} which upon multiplication by $\frac{1}{2}\eta^{kmn}$, yields
\begin{align}
- \frac{1}{2} \eta^{kmn}(\partial_n F_{0m} -  \partial_m F_{0n}) =  \frac{1}{2} \eta^{kmn}\partial_0 F_{mn}\cr
 \frac{1}{2} \eta^{kmn}(\partial_m F_{0n} -  \partial_n F_{0m}) = \frac{1}{\sqrt{\gamma}}\partial_0 \{\sqrt{\gamma}~(\frac{1}{2}\eta^{kmn}F_{mn}~)\},
\end{align}
where we used the fact that $F_{n0} = -F_{0n}$. Employing the  definition of curl \eqref{curl}, suggests the following definitions of the electric and magnetic fields,
\begin{equation}
E_i = F_{0i} \;\;\; ; \;\;\; B^{i}=-\frac{1}{2}\eta^{ijk}F_{jk} \label{3}
\end{equation}
so that it reduces to the familiar form
\begin{equation}
\nabla \times {\bf E} = - \frac{1}{c\sqrt{\gamma}}~\frac{\partial}{\partial{t}}(\sqrt{\gamma}~ {\bf B})\label{ME3}
\end{equation}
which is the equivalent of the same equation in flat spacetime in a curvilinear coordinate. Applying the same procedure, and using the  definition of divergence \eqref{div}, equation \eqref{CME11} reduces to
\begin{equation}
\nabla . {\bf B} \equiv \frac{1}{\sqrt{\gamma}}~\frac{\partial}{\partial{x^i}}(\sqrt{\gamma}~B^i)=0  \label{ME1}
\end{equation}
To write the second pair of Maxwell equations in 3D form, first we need to express the current 4-vector, and its components in the threading decomposition formalism. This is given by \cite{LL},
\begin{equation}
j^\mu = \frac{\rho c}{\sqrt{g_{00}}}\frac{dx^\mu}{dx^0} \equiv \frac{1}{\sqrt{g_{00}}} (\rho c, {\bf j}),
\end{equation}
so that  the second pair of Maxwell equations are obtained by decomposing the equation \eqref{CME2} into its space-time components as follows
\begin{gather}
-\frac{1}{\sqrt{\gamma}}\partial_i (\sqrt{-g} F^{0i})= \frac{1}{\sqrt{\gamma}}\partial_i \{\sqrt{\gamma}(-\sqrt{g_{00}}F^{0i})\} =  {4\pi}\rho \label{CME13}\\
\frac{1}{\sqrt{\gamma}}\partial_\mu (\sqrt{-g} F^{i\mu})=-\frac{1}{\sqrt{\gamma}}\{\partial_0 (\sqrt{\gamma} (-\sqrt{g_{00}}F^{i0})) - \partial_j (\sqrt{\gamma} (-\sqrt{g_{00}} F^{ij})\} = - \frac{4\pi}{c}j^i, \label{CME14}
\end{gather}
which again in comparison to their counterparts in flat spacetime in curvilinear coordinates, suggest the  following definitions of the displacement current $\bf D$ and the auxiliary magnetic field $\bf H$ \footnote{As pointed out previously, one can show that the definitions of $E_i$ and $H_i$ given above could be obtained in terms of the projection of the field tensor along a given 4-vector. For example here one can show that the 4-velocity $u^{\alpha} = (1,0,0,0)$ of any observer in their comoving coordinate system could be used to define these fields, i.e,
\begin{equation}
E_{\mu} = -F_{\mu \nu}u^{\nu} = F_{0\mu}\;\;\; ; \;\;\;
H_{\mu}=-\frac{1}{2} \eta_{\mu \nu \alpha \beta}u^\nu F^{\alpha \beta}. \label{H}
\end{equation}
The above 4-vectors are equivalent to the 3-vectors defined in \eqref{3}, and \eqref{DHT}, since here $E_0 = H_0 = 0$.  In the case of a {\it stationary} spacetime this is equivalent to the projection along the  timelike Killing vector field of the spacetime.},
\begin{equation}\label{DHT}
D^{i}=-\sqrt{g_{00}}F^{0i} \;\; ; \;\;  H_{i}=-\frac{1}{2}\sqrt{g_{00}} \eta_{ijk}F^{jk}\;\;  \left( F^{ij}=-\frac{1}{\sqrt{g_{00}}} \eta^{ijk} H_k \right).
\end{equation}
Substituted in \eqref{CME13}, and \eqref{CME14}, we find them, as expected, in the following 3D forms,
\begin{gather}
\nabla . {\bf D} \equiv \frac{1}{\sqrt{\gamma}}~\frac{\partial}{\partial{x^i}}(\sqrt{\gamma}~D^i) = 4 \pi \rho \label{ME2} \\
\nabla \times {\bf H} = \frac{1}{c\sqrt{\gamma}}~\frac{\partial}{\partial{t}}(\sqrt{\gamma}~{\bf D}) + \frac{4\pi}{c} {\bf j}. \label{ME4}
\end{gather}
The above definitions \eqref{3},and \eqref{DHT} of the electromagnetic fields in terms of the electromagnetic field tensor show very clearly that these fields are not independent, and indeed one can obtain their relation by rewriting  the spatial and space-time components of the field tensor, namely $F_{0i} = g_{0\mu}g_{i\nu}F^{\mu\nu}$, and $ F^{ij} = g^{i\mu}g^{j\nu}F_{\mu\nu}$, in terms of these fields. It is not difficult to show that this leads to the following {\it gravitationally-induced constitutive equations} in a given spacetime
\begin{equation}\label{ce}
{\bf D}=\frac{{\bf E}}{\sqrt{g_{00}}} + {\bf H} \times {\bf g}~~~~;~~~~ {\bf B}=\frac{\bf H}{\sqrt{g_{00}}} - {\bf E} \times {\bf g}.
\end{equation}
The above equations are very similar to the constitutive equations for an accelerated observer in a matter-free space such as the one in a rotating frame of reference \cite{Heer}. This is obviously expected from the equivalence principle. Two points need to be stressed here: \\
{\bf 1)}- In the above equations the vacuum permittivity $\epsilon_0$ and permeability $\mu_0$ do not appear since the Gaussian units are used. \\ {\bf  2)}- It is the magnetic induction $\bf B$ which  is defined to be divergence-free, while the {\it auxiliary} magnetic field $\bf H$ is not necessarily divergence-free.
%%%%%%%%%%%%%%%%%%%%%%%%%%%%%%%%%%%%%
\section{Spacetime index of refraction from constitutive equations}
For static spacetime the constitutive equations \eqref{ce} reduce to
\begin{equation}\label{ce1}
{\bf D}=\frac{{\epsilon_0\bf E}}{\sqrt{g_{00}}}=\epsilon {\bf E} ~~~~;~~~~ {\bf B}=\frac{\mu_0 \bf H}{\sqrt{g_{00}}}= \mu {\bf H}.
\end{equation}
in which, for comparison purpose, we have reintroduced the vacuum electric permitivity and magnetic permeability $\epsilon_0$ and $\mu_0$ respectively (i.e employing momentarily non-Gaussian units).
Comparing these equations with their counterparts in flat spacetime, namely ${\bf D} = \epsilon_0 {\bf E}$, and ${\bf B} = \mu_0 {\bf H}$,  and the fact that in these units $c=\frac {1}{\sqrt{\epsilon_0 \mu_0}}$, we have
\begin{equation}
{\tilde c} = \frac {1}{\sqrt{\epsilon \mu}}=c {\sqrt{g_{00}}}
\end{equation}
as the velocity of light in a static spacetime. In other words one can interpret $\frac {1}{\sqrt{g_{00}}}$ as  the index of refraction for static spacetimes, which is obviously in agreement with our previous definitions in section III.\\
To show the consistency of the above definitions of electromagnetic fields in a curved spacetime, and specifically the definition of the spacetime index of refraction for stationary spacetimes, namely \eqref{IRS2} and \eqref{Index2}, here we aim at finding the same definition from the constitutive equations \eqref{ce}.\\
To do so we start from the previously mentioned fact that in geometric optics, propagation of light in empty regions of a curved spacetime is locally approximated by monochromatic plane waves \cite{MTW}. In other words, in this limit,  a vacuum gravitational field acts on light as a non-dispersive medium. Indeed this is the working principle both in gravitational lensing studies \cite{ShEF}, as well as in the studies related to the so called singularity theorems in black hole physics \cite{HE}. So all our fields have the following plane wave form
\begin{equation}\label{PW}
 ({\bf E}, {\bf D}, {\bf B}, {\bf H}) \propto e^{-i{k_\mu} {x^\mu}} = e^{-i(\omega_0 t + {k_i}{x^i} )}.
\end{equation}
in which we note that $k_i$ is given by \eqref{kd}. On the other hand in a stationary spacetime, the two Maxwell equations  \eqref{ME3}, and \eqref{ME4} in the absence of sources reduce to
\begin{gather}
\nabla \times {\bf E} = - \frac{1}{c}~\frac{\partial}{\partial{t}} ~{\bf B} \label{MEF1}\\
\nabla \times {\bf H} = \frac{1}{c}~\frac{\partial}{\partial{t}} ~{\bf D}. \label{MEF2}
\end{gather}
Now to apply the general plane wave form \eqref{PW} in the above equations,
we need to calculate the curls of $\bf E$, and $\bf H$. Calculations are similar, so we only give the details for \eqref{MEF2}, starting with its left hand side,
\begin{equation}
(\mathbf{\nabla}\times\mathbf{H})^{i} = \frac{1}{2\sqrt{\gamma}} \epsilon^{ijk}(\partial_{j}H_{k}-\partial_{k}H_{j})
=\frac{1}{\sqrt{\gamma}} \epsilon^{ijk}\partial_{j}H_{k},\label{a2}
\end{equation}
Now substituting the plane wave forms of $\bf H$, and $\bf D $ in the above equation, and in the right hand side of \eqref{MEF2} respectively, we have
\begin{align}
(\mathbf{\nabla}\times\mathbf{H})^{i}&=\frac{-i}{\sqrt{\gamma}}\epsilon^{ijk} H_{k}k_{j} =\frac{i}{\sqrt{\gamma}} \epsilon^{ijk}H_{k}\tilde{k}_{j}+\frac{i~\omega_{0}}{c\sqrt{\gamma}} \epsilon^{ijk}H_{k} g_{j}
\\ &=\frac{-i\omega_{0}}{c}D^{i},\label{a3}
\end{align}
where in the first line we used the important relation \eqref{kd}, and denoted the covariant counterpart of $k^i$ (a covariant vector in $\Sigma_3$) by ${\tilde k}_i = \gamma_{ij}k^j$. Using the definition of the cross product in $\Sigma_3$ we can write the above equation in the following vector form ,
\begin{equation}
{\bf D} =  \frac{c}{\omega_0} ({\bf H} \times {\bf k}) + ({\bf H} \times {\bf g}),
\end{equation}
in which, to keep the notation consistent, we have  $({\bf k})_i \equiv {\tilde k}_i$, i.e the components of the covariant vector ${\bf k}$ in $\Sigma_3$ are denoted by ${\tilde k}_i$. Also using the same procedure, one can show that the equation \eqref{MEF1}
with the plane wave form for the $\bf E$ and $\bf B$ fields, leads to the following relation
\begin{equation}
{\bf B} =  \frac{c}{\omega_0} ({\bf k} \times {\bf E})- ({\bf g} \times {\bf E}).
\end{equation}
‌By substituting the above relations for $\bf D$ and $\bf B$ in Eqs. \eqref{ce} we find
\begin{gather}
{\bf E} = \frac{\sqrt{g_{00}}{c}}{\omega_0} ({\bf H} \times {\bf k}) = {\bf H} \times {{\bf {\hat k}}} \label{} \\
{\bf H} = - \frac{\sqrt{g_{00}}{c}}{\omega_0} ({\bf E} \times {\bf k}) = - {\bf E} \times {{\bf {\hat k}}} \label{}
\end{gather}
where we have used \eqref{dk2}. The above equations show that the three vectors $\bf E, \bf H$, and $\bf {\hat k}$ are mutually orthogonal. Now using the above relations for $\bf E$, and $\bf H$ in the second and  first constitutive equations \eqref{ce} respectively, we find
\begin{align}
\mathbf{D}=\frac{\mathbf{E}}{\sqrt{g_{00}}}+\mathbf{g} \times (\mathbf{E}\times \hat{\bf k}),\\
\mathbf{B}=\frac{\mathbf{H}}{\sqrt{g_{00}}}+\mathbf{g}\times(\mathbf{H}\times\hat{\bf k}).
\end{align}
Applying the relation $
\mathbf{a}\times(\mathbf{b}\times \mathbf{c})=\mathbf{b}(\mathbf{a}.\mathbf{c})- \mathbf{c}(\mathbf{a}.\mathbf{b})$
to the above equations we end up with the following equations,
\begin{align}
{\bf D}=\left\lbrace  \frac{1}{\sqrt{g_{00}}}+{\bf g}.{\hat{\bf k}}\right\rbrace {\bf E} - {\hat{\bf k}}({\bf g}.{\bf E}) = \epsilon{\bf E} -{\hat{\bf k}}({\bf g}.{\bf E}) \label{CE1} \\
{\bf B}=\left\lbrace  \frac{1}{\sqrt{g_{00}}}+{\bf g}.{\hat{\bf k}}\right\rbrace {\bf H} - {\hat{\bf k}}({\bf g}.{\bf H}) = \mu {\bf H} -{\hat{\bf k}}({\bf g}.{\bf H}), \label{CE2}
\end{align}
in which comparing with their electromagnetic counterparts, and using the fact that the second terms in the right hand side of the above equations are along $\bf {\hat k}$ which  is orthogonal to both $\bf E$, and $\bf H$, we have  introduced the electric permittivity $\epsilon$, and magnetic permeability $\mu$ of a stationary spacetime as
\begin{equation}
\epsilon = \mu = \frac{1}{\sqrt{g_{00}}}+{\bf g}.{\hat{\bf k}},
\end{equation}
so that the spacetime index of refraction is given by
\begin{equation}\label{N}
n_T = {\sqrt{\epsilon \mu}} =  \frac{1}{\sqrt{g_{00}}}+{\bf g}.{\hat{\bf k}}.
\end{equation}
It was noticed that the basis of this analogy is provided by the application of the geometric optics assumption in curved spacetimes.
%%%%%%%%%%%%%%%%%%%%%%%%%%%%%%%%%%%%%%%%%%%%%%%%%%%%%%%%%%
\section{spacetime index of refraction in slicing formalism}
Here we show that as in the case of threading decomposition formalism, one can define spacetime index of refraction in slicing formalism using three different approaches. In the next two subsections we use Fermat's principle and classical definition of refractive index to arrive at such a definition. In the next section we will employ the gravitationally-induced constitutive equations to find the same definition.
%%%%%%%%%%%%%%%%%%%%%%%%%%%%%%%%%%%%%%
\subsection{Spacetime index of refraction from Fermat's principle}
To define the  spacetime refractive index in slicing formalism,  we start from the Fermat's principle in its general form  \eqref{FP-2},
\begin{equation}\label{FP22}
\delta \int  k^0 (\frac{d\lambda}{dl})~dl = 0,
\end{equation}
and try to write the integrand in terms of the quantities defined in the slicing formalism. To this end, for a monochromatic wave with $k_{0}=\frac{\omega_{0}}{c}$, we use the relation $ds^2=0$ in \eqref{metric2}, so that
\begin{equation}
d l_S =  N dx^0 \;\;\; \Longrightarrow \;\;\; \frac{dl_S}{d\lambda} = N k^0
\end{equation}
Substituting the above relation in \eqref{FP22}, we find in a remarkably simple way the desired result,
\begin{equation}\label{FP221}
\delta \int \frac{dl_S}{N} = \delta \int n_S~{dl_S} = 0,
\end{equation}
Showing that in slicing formalism, the spacetime index of refraction is given by
\begin{equation}\label{n22}
n_S  = \frac{1}{N} = \sqrt{g^{00}}.
\end{equation}
%%%%%%%%%%%%%%%%%%%%%%%%%%%%%%%%%
\subsection{Spacetime index of refraction from classical definition}
Employing the general  relation \eqref{Index} for the slicing formalism we have
\begin{align}\label{irs3}
n_S=\frac{dx^{0}}{dl_S}.
\end{align}
Now setting $ds^2 = 0$ in \eqref{metric2} we find
\begin{equation}
dl_S = N dx^0,
\end{equation}
which upon substitution in \eqref{irs3} leads to
\begin{equation}
n_S  = \frac{1}{N} = \sqrt{g^{00}}.
\end{equation}
Incidentally, using the general relation \eqref{Index} in Fermat's principle, it takes the following form of an action principle,
\begin{equation}\label{FPS}
\delta \int dx^0 \equiv \delta S =  0.
\end{equation}
With the integral taken over the spatial paths in $\Sigma_3$
(or $\Sigma_t$), the null rays are nothing but the extremals of the above action. This is indeed the form of the Fermat's principle as the variation of the time of arrival, applied to a  stationary spacetime, which was introduced by Kovner \cite{Kovner}, and later proved by Perlick \cite{Perlick} for general spacetimes. By the above argument, in the case of stationary spacetimes, one can decompose the spacetime metric into temporal and spatial sectors, which are then related for null curves by setting $ds^2 = 0$. Now the null rays are equivalently characterized, either through the above  temporal variational principle, or the spatial variational principle  $\delta \int n~dl =  0$, with the corresponding spatial line element, and the spacetime index of refraction introduced in the decomposition formalism.
%%%%%%%%%%%%%%%%%%%%%%%%%%%%%%%%%%%%%%%%%
\section{Maxwell equations and gravitationally-induced constitutive equations in slicing decomposition formalism}
To find gravitationally-induced constitutive equations in slicing formalism, first we derive Maxwell equations in their 3D form in this formalism. Here, as in the threading formalism, we define the fields in such a way that one arrives at the Maxwell equations which are similar in form to their counterparts in flat space in curvilinear coordinates. To this end we start with
the Maxwell equations in terms of the  electromagnetic tensor given in  \eqref{CME1}-\eqref{CME2}, and  employ the same procedure as we did in the case of threading decomposition in section IV. \\
Since the Levi-Civita pseudotensor in this formalism is basically of the same form as in threading formalism (refer to appendix), starting from the spatial components of the source-free equation \eqref{CME1},  we arrive at the same equations \eqref{ME3}, and \eqref{ME1}, with similar definitions for the fields $\bf E$, and $\bf B$ namely,
\begin{equation}\label{BES}
	B^{i}=-\frac{1}{2} \eta^{ijk} F_{jk} \;\;\;  ; \;\;\;
	E_{i}=F_{0i}.
\end{equation}
For the second pair of Maxwell equations, in analogy with the threading formalism,  we define the current 4-vector as
\begin{equation}
j^{\mu}=\frac{1}{N}\left( \rho c, \mathbf{j}\right)
\end{equation}
Now to arrive at equations similar in form to their flat space counterparts in curvilinear coordinates, i.e, \eqref{ME2}, and \eqref{ME4}, we need to define the fields $\bf D$, and $\bf H$ as follows
\begin{equation}\label{DHS}
D^{i} = -N F^{0i} \;\;\; ; \;\;\;
H_{i} = - \frac{N}{2} \eta_{ijk} F^{jk},
\end{equation}
So in both decomposition formalisms,  we arrive at  equations similar in form  to their flat spacetime counterparts, if we define the fields $\bf E$, and $\bf B$, according to Eqs. \eqref{BES} (similar in form in both formalisms), and  the fields $\bf D$, and $\bf H$, according to \eqref{DHT} in the  threading formalism, and according to \eqref{DHS} in slicing formalism.\\
As in the case of threading formalism, the above-introduced  electromagnetic fields are not independent, and their relation constitute the gravitationally-induced  constitutive equations in slicing decomposition formalism. Indeed starting from the following relations
\begin{equation}
 F^{0i} = g^{0\mu}g^{i\nu} F_{\mu\nu}\;\;\; ; \;\;\;
 F_{ij} = g_{i\mu}g_{j\nu} F^{\mu\nu},
\end{equation}
by writing the components of the electromagnetic field tensor in terms of the introduced 3D electromagnetic fields, and the metric elements in terms of the lapse function and shift vector, one arrives at the following constitutive equations
\begin{equation}\label{ces}
{\bf D}=\frac{{\bf E}}{N} + \frac{1}{N}{\bf B} \times {\bf N}~~~~;~~~~ {\bf B}=\frac{\bf H}{N} - \frac{1}{N}{\bf D} \times {\bf N}.
\end{equation}
%%%%%%%%%%%%%%%%%%%%%%%%%%%%%%%%%%%%%%%%
\subsection{Spacetime index of refraction from constitutive equations}
Looking at the definition of the spacetime index of refraction in slicing formalism obtained in previous sections, namely $n_S$, it is obvious that unlike its definition in threading formalism, there is no trace of the wave vector in its expression. It is purely a geometric definition. This is a hint that we do not need to implement plane wave form of fields in equations \eqref{ces} to arrive at this definition. Indeed, since all the electromagnetic fields are now defined on $\Sigma_t$, and the shift vector is tangent to this hypersurface, the second terms in the right hand side of the gravitationally-induced constitutive equations \eqref{ces}, are orthogonal to it. So we can write equations \eqref{ces} as follows
\begin{equation}\label{cesi}
{\bf D}={\epsilon}{\bf E} + \frac{1}{N}{\bf B} \times {\bf N}~~~~;~~~~ {\bf B}=\mu {\bf H} - \frac{1}{N}{\bf D} \times {\bf N}.
\end{equation}
in other words now we have
\begin{equation}\label{sirce}
\epsilon = \mu = \frac{1}{N}
\end{equation}
so that the spacetime index of refraction is given by
\begin{equation}\label{N}
n_S = {\sqrt{\epsilon \mu}} = \frac{1}{N} = \sqrt{g^{00}}.
\end{equation}
%%%%%%%%%%%%%%%%%%%%%%%%%%%%%%%%%%%%%%%%%%%%%%
\section{Comparison between the two indices of refraction}
By using the spacetime index of refraction one could in principle calculate the light ray trajectories in the corresponding spacetime, or its metamaterial analog \cite{NPF,PFRN}. These trajectories, on the other hand are independent of the decomposition formalism, and the related observer. So on physical grounds it is expected that one could transform the above two definitions of the spacetime refractive indices into one another, and this is indeed a consistency check of the above calculations. Here, starting from the Fermat's principle in slicing formalism \eqref{FP221}, we show that one arrives at its counterpart in threading formalism \eqref{IRS2}.\\
To do so we start from the 3D line element in slicing decomposition formalism,
\begin{align}\label{equiv}
	dl^{2}_{S}&=-g_{ij}(dx^{i}+N^{i}dx^{0})(dx^{j}+N^{j}dx^{0}) \nonumber \\
	&=-g_{ij}~dx^{i}dx^{j}-g_{ij}N^{i}N^{j}(dx^{0})^2-2~g_{ij} N^{i}dx^{j}dx^{0} \nonumber \\
	&=	dl^{2}_{T}-g_{00}g_{i}g_{j}dx^{i}dx^{j}-g_{ij}N^{i}N^{j}(dx^{0})^2-2~g_{ij} N^{i}dx^{j}dx^{0}
\end{align}
where in the third line we used the definition of $dl_T$ in \eqref{dl}. Now the right hand side of the above equation could be written in the following form
\begin{align}
dl^{2}_{S}&=	dl^{2}_T \left( 1-g_{00}~g_{i}g_{j}\frac{dx^{i}}{dl_T}\frac{dx^{j}}{dl_T}-g_{ij}N^{i}N^{j}(\frac{dx^{0}}{dl_T})^2-2~g_{ij} N^{i}\frac{dx^{j}}{dl_T}\frac{dx^{0}}{dl_T}\right).
\end{align}
Employing the relations \eqref{dk2} and \eqref{Index} for $\hat{k}^{i}$, and $n_T$ respectively, as well as the first equation in \eqref{efs},  we find
\begin{align}
dl^{2}_{S}&=	dl_T^{2} \left( 1-g_{00}~\left(\mathbf{g}.\hat{k} \right)^{2} + n_T^2 (N^2-g_{00})+2N_{j}\hat{k}^{j} n_T\right) \nonumber \\
&=	dl_T^{2} \left( 1-g_{00}~\left(\mathbf{g}.\hat{k} \right)^{2} + n_          T^2(N^2-g_{00}) + 2n_T g_{00} \mathbf{g}.\hat{k}\right) \nonumber \\
&=	dl_T^{2} \left( 1-g_{00}~\left(n_T-\mathbf{g}.\hat{k}\right)^{2}+n_T^2N^2\right).
\end{align}
Now using the equation \eqref{IRS2} we end up with
\begin{align}
\frac{1}{N}dl_{S} = n_T~dl_{T}.
\end{align}
Apart from the above direct approach, one could have obtained this result (though not with the exact form of the index of refraction in the threading formalism) from writing Eq.\eqref{Index} for the two decomposition formalisms, and equating the coordinate time differential according to the relation \eqref{FPS}. This shows the consistency of the two definitions introduced for the spacetime index of refraction in the two decomposition formalisms.\\
Now the question arises that in simulating ray trajectories in a stationary spacetime, using the spacetime index of refraction in the context of Fermat's principle, which decomposition formalism
should be used?. Of course this is a question of implementation because as shown above they are equivalent, derived from the same general relation \eqref{FP}. This question is specifically of importance noting that in threading approach the refractive index depends both on the position, and {\it direction} of the ray, whereas its definition in slicing approach, as pointed out previously, is purely geometrical and does not depend on ray's direction. This fact is obviously misleading, since apart from the refractive index we also need to know the spatial line element to implement Fermat's principle. Now in this regard, as pointed out in Remark I in section IV, it is the spatial line element of threading approach which is proportional to the affine parameter of null geodesics, and naturally adapted to ray tracing simulations.\\
Indeed in implementing the spatial line element in slicing approach,  one can not use the complicated form \eqref{dlf}, from which we started to show the consistency of the two definitions in \eqref{equiv}. So in looking for the application of the spacetime index of refraction in the next section, we will use the threading decomposition formalism.
%%%%%%%%%%%%%%%%%%%%%%%%%%%%%%%%%%%%%%%%%%%%%%%%%%%%%%%%%%%
\section{Application}
The most obvious application of the spacetime index of refraction is in the light bending, and gravitational lensing studies in which, as mentioned before, the geometric optics limit is the working assumption. But obviously one should be cautious with the fact that it is defined in the spatial sector of the spacetime in a decomposition formalism corresponding to an observer. For this reason, using this approach is perhaps more effective in ray tracing simulations in the corresponding studies. An example could be the study of the effect of cosmological constant in light bending in the context of Scwharzschild-de Sitter spacetime which is still a controversial subject \cite{Ishak1}-\cite{Simpson}. Indeed one can use the above ideas to simulate light paths in Schwarzschild-de Sitter spacetime, and compare them with the corresponding paths in Schwarzschild spacetime to see the effect of the cosmological constant \cite{FMMNP}.\\
As another interesting example in stationary spacetimes, is the possibility of nonreciprocal refraction, and light paths due to the fact that now the index of refraction depends on both the grvitomagnetic vector potential, and the direction of the ray propagation, in the form $\bf g.\hat {\bf k}$ in threading decomposition formalism. Nonreciprocal refraction could happen  if the light rays pass in opposite directions through a domain wall separating  two neighbouring spacetime regions with opposite  gravitomagnetic vector potentials $\bf g$.
This could be taken as the gravitational analog of nonreciprocal refraction through a toroidal domain wall in which the index of refraction has the same property of depending on the direction of ray propagation as well as on toroidal moment \cite{Sawada}-\cite{BMN}. Indeed light rays in NUT spacetime could provide such a scenario. Metric of this spacetime  in Schwarzschild-type ($t, r, \theta, \phi$) coordinates is given by
\begin{equation}\label{n1}
ds^2 = f(r) (dt - 2l \cos\theta d\phi)^2 - \frac{dr^2}{f(r)} - (r^2 + l^2) d\Omega^2
\end{equation}
with
\begin{equation}\label{f}
f(r) \equiv g_{00}=\frac{r^2 - 2mr - l^2}{r^2 + l^2},
\end{equation}
in which $m$, and  $l$ are the mass and  NUT (magnetic mass) parameters respectively. So, compared to metric form \eqref{metric1}, it has the gravitomagnetic vector potential
\begin{equation}\label{NGVP}
 {\bf g} = (0, 0, g_\phi) \equiv (0, 0, 2l \cos\theta),
\end{equation}
which is obviously zero in the equatorial plane and positive in the northern hemisphere, and negative in southern hemisphere, so that the hypersurface $\theta = \frac{\pi} {2}$ acts as the garvitational analog of a toroidal domain wall. Now it is expected that a light ray with an incoming angle $\theta = \frac{\pi} {2} - \delta$ towards this hypersurface follows a different path from that of an outgoing light ray crossing this plane in the opposite direction. \\
Another example in the same context is the nonreciprocal light paths in the equatorial plane of Kerr geometry. It is noted that the same effect is not expected to be present in the equatorial NUT plane where the  gravitomagnetic vector potential \eqref{NGVP} vanishes. In the case of the kerr spacetime, on physical grounds the dragging effect in the equatorial plane will have different effects on prograde and retrograde null rays, leading to nonreciprocal paths, which is expected to be stronger as we get closer to the horizon. Applying the effect near the horizon one should be careful by taking into account the second condition for the applicability of the geometric optics approximation, namely the smallness of the wavelength with respect to spacetime radius of curvature. The above effects will be considered elsewhere \cite{MEN}, but in what follows we will study an interesting gravitational analog of a toroidal moment in the weak field approximation of stationary spacetimes.
%%%%%%%%%%%%%%%%%%%%%%%%%%%%%%%%%%%%%%%%%%%%%%%%%%%%%
\subsection{Gravitational analog of a toroidal moment}
There is an interesting analogy between the spacetime index of refraction \eqref{IRS2}, and the refractive index of multiferroic materials characterized by a toroidal moment \cite{Sawada},
\begin{equation}
n({\bf r}) = n_{0} ({\bf r})+ \alpha {\bf \hat{k}} \cdot {\bf T}(\bf{r}),
\label{105}
\end{equation}
where
\begin{equation}\label{Toro}
{\bf T} \propto \int {\bf r} \times {\bf S}({\bf r})\; d^3{\bf r},
\end{equation}
with ${\bf S}(r)$ the spin density \cite{toroidal}. In its gravitational analog \eqref{IRS2}, $n_0$ corresponds to $ \frac{1}{\sqrt{g_{00}}} $, and $\bf {g}$  behaves analogously to the toroidal moment $\bf { T}$. Indeed, the presence of the toroidal moment, and whether it is parallel or antiparallel to the direction of propagation ${\bf \hat{k}}$ induces the so called optical magnetoelectric effect (OME), and the nonreciprocal refraction from a toroidal domain wall \cite{Sawada}.
Building on the above-mentioned analogy, we show that there is a gravitational analog of relation \eqref{Toro} between  $\bf g$ and the gravitational angular momentum density of source particles in the weak-field approximation of these spacetimes. This in turn will show that there should be a gravitational analog of nonreciprocal refraction caused by the presence of the  gravitomagnetic potential $\bf g$ in the refractive index of stationary spacetimes. To this end we begin with Fermat’s principle in the threading decomposition formalism, Eq. \eqref{FP-4}, in the following form,
\begin{equation}
\delta \int n_T \frac{dl_T}{d\lambda} d\lambda = \delta \int n_T \sqrt{\gamma_{ij} \dot{x}^i \dot{x}^j}  d\lambda = 0,
\label{102}
\end{equation}
where, as previously noted, $n_T=\frac{1}{\sqrt{g_{00}}}+{\bf g}.{\bf \hat{k}}$, and $\dot{x}^i=\frac{dx^i}{d\lambda}= k^i$. It is expected that Eq. \eqref{102}  leads to the light trajectory in agreement with the spatial component of the null geodesic equation in the threading decomposition formalism, as a force equation on light rays. Indeed the above variation leads to the following Euler-Lagrange equation,
\begin{equation}
\frac{d}{d\lambda} (n_T \gamma_{ij} \hat{k}^j) =  \partial_{i} (  n_T \sqrt{\gamma_{ij} \dot{x}^i \dot{x}^j}).
\label{103}
\end{equation}
Restating the above equation in terms of the differentiation with respect to $dl_T$, we end up with,
\begin{equation}
\frac{{\cal D}\left(n_T \hat{k}^i \right)}{{\cal D} l_T} \equiv \frac{d}{dl_T} \left(n_T \hat{k}^i \right)+  n_T \lambda^{i}_{j m} \hat{k}^j \hat{k}^m = \gamma^{ij} \partial_j n_T,
\label{104}
\end{equation}
where $\frac{\cal D}{{\cal D} l_T}$ is the 3D covariant derivative along the ray, defined  in $\Sigma_3$ in terms of $\gamma_{ij}$, and the 3D Christoffel symbols ($\lambda^{i}_{j m}$) obtained from its derivatives \cite{LL}. This equation is similar in form to the equation of light rays in geometric optics limit in a medium with a refractive index $n({r})$ \cite{Born}, namely,
\begin{equation}
\frac{d}{ds} \left(n({\bf{r}}) \frac{d{\bf r}}{ds}\right) = {\bf \nabla}n({\bf{r}}).
\end{equation}
which, itself is similar to the force equation in particle mechanics. Indeed one can interpret equation \eqref{104} as a gravitational force equation acting on light rays distorting the light trajectory in $\Sigma_3$ and forcing them to follow the null geodesics in 4D spacetime.
Now substituting for the refractive index $n_T$ we find,
\begin{equation}
\frac{{\cal D}\left(n_T \hat{k}^i \right)}{{\cal D} l_T} = \gamma^{ij} \partial_j n_T =  \gamma^{ij} \left({\partial}_j ( \frac{1}{\sqrt{g_{00}}} ) + ({\bf \hat{k}} \cdot {{\bf \nabla}}){g}_j
+ ({\bf \hat {k}} \times ({\bf {\nabla}} \times {\bf {g}})_j \right),
\label{105}
\end{equation}
which is similar in form to equation (3) in \cite{Sawada}. As expected, the first and third  terms on the right-hand side of the above equation include the Lorentz-type force terms, with the third term representing the effect of the gravitomagnetic field ${\bf B}_{g} = {\bf \nabla} \times {\bf g}$ on light rays \cite{LBNZ}, \cite{FMMNP}. On the other hand  the ${\bf {\hat k}}$-dependence in the last two terms leads to the gravitational analog of the nonreciprocal refraction in materials with a toroidal domain wall \cite{Sawada}.\\
To illustrate the above analogy more clearly, we find the analog of the relation \eqref{Toro} in the weak-field limit of the Einstein field equations with the linearized equations,
\begin{equation} \label{106}
\Box h_{\mu\nu} - 2 \partial_{(\mu} \partial_\alpha h^{\alpha}_{\nu)} + \partial_\mu \partial_\nu h - \eta_{\mu\nu} \Box h + \eta_{\mu\nu} \partial_\alpha \partial_\beta h^{\alpha\beta} = -16\pi T_{\mu\nu},
\end{equation}
where $ g_{\mu \nu} = \eta_{\mu \nu} + h_{\mu \nu} $ and $ |h_{\mu\nu}| << 1$ are assumed. Rewriting the above equation in the Hilbert-Einstein gauge, $\partial_\mu \left( h^{\mu \nu} - \frac{1}{2} h \eta^{\mu \nu} \right) = 0$, for stationary fields, and sources (i.e $ h^{\mu \nu},_0 = 0 ,  T^{\mu \nu},_0 = 0 $), we have,
\begin{equation} \label{107}
\nabla^2 ( h^{\mu \nu} - \frac{1}{2} \eta^{\mu \nu} h)= - 16 \pi T^{\mu\nu}
\end{equation}
 The solution to the Poisson equation \eqref{107} is given by,
\begin{equation} \label{108}
\bar{h}^{\mu\nu} \equiv h^{\mu \nu} - \frac{1}{2} h \eta^{\mu \nu}= - 4 \int \frac{T^{\mu\nu}(\mathbf{r}')}{|\mathbf{r} - \mathbf{r}'|} \, d^3 {\bf r}'.
\end{equation}
On the other hand it is not difficult to show that the gravitoelectromagnetic Lorentz-type force \cite{LBNZ}, acting on a particle  in the weak-field, and low velocity limit of stationary spacetimes takes the following form
\begin{equation} \label{109}
{\bf f}_g = - {\bf {\nabla}}(\frac{1}{2} h_{00}) + {\bf {v}}
\times {\bf {B}}_g,
\end{equation}
where  ${\bf {B}}_g = {\bf \nabla} \times {\bf {g}}$, with $ g_i = -h_{i0} = -\bar{h}_{i0} = \bar{h}^{i0} $.
Now to find an expression for $\bf {g}$ in terms of the integrals over components of the energy momentum tensor,  we consider \eqref{108} under the dipole approximation, namely,
\begin{equation} \label{110}
\bar{h}^{\mu\nu} \sim -\frac{4}{r} \int T^{\mu\nu}(\mathbf{r}') \, d^3 {\bf r}' - \frac{4}{r^3} x^k \int x'_k T^{\mu\nu}(\mathbf{r}') \, d^3 {\bf r}'.
\end{equation}
Using the conservation of the energy-momentum tensor in the weak field limit, and after some mathematical manipulation, the components of $ \bar{h}^{\mu \nu} $ can be computed from \eqref{110} as follows:
\begin{equation}
\begin{cases}
\bar{h}^{00} = -4 (\frac{M}{r} +
\frac{{\bf {p}} \cdot {\bf {r}}}{r^3}) \\
\bar{h}^{i0} = \frac{2}{ r^3} \left( {\bf r} \times {\bf L}\right)^i   \\
\bar{h}^{ij} = 0
\end{cases}
\label{111}
\end{equation}
in which we have used the following definitions;
\begin{equation}
\begin{cases}
M = \int T^{00} d^3 {\bf r}' \\
{\bf {P}}_g = \int {\bf {r'}} \, T^{00} \, d^3 {\bf r}' \\
{\bf  L} = \int {\bf {r'}} \times {{\bf {\cal P}}} \, \, d^3 {\bf r}' \\
{\cal P}^i=T^{i0}
\end{cases}
\label{112}
\end{equation}
Here, $ {\cal P}^i $ represents the momentum density of source particles across a spacelike hypersurface, and consequently, the total angular momentum of source particles is given by ${\bf L} = \int {\bf r} \times {\bf {\cal P}} \, d^3{\bf r}$. Also ${\bf P}_g$ is interpreted as the gravitational analog of the electric dipole moment.
Indeed the second equation in  \eqref{111}, shows that the gravitomagnetic vector potential in the weak-field, dipole approximation is given by,
\begin{equation} \label{113}
{\bf g} = \frac{2}{r^3}({\bf r} \times {\bf L}).
\end{equation}
Defining ${\bf {\cal L}} = {\bf r} \times {\bf {\cal P}}$ as the density of the angular momentum of the source particles, from the above relation we have,
\begin{equation}
{\bf g} \sim \int {\bf r}  \times {\bf \cal{L}}({\bf r}')  \, d^3{\bf r}',
\end{equation}
which is the gravitational analog of \eqref{Toro}. In other words in the weak-field, dipole approximation of Einstein field equations for stationary sources and field, the gravitomagnetic vector potential $ {\bf g}$ behaves analogously to a toroidal moment \cite{toroidal}.\\
By the above analogy it is expected that there should be nonreciprocal refraction for light rays propagating in opposite directions in a region of spacetime with a gravitational analog of a toroidal domain wall. This could happen, for example, in the asymptotically flat region of the NUT spacetime, which as mentioned previously, has a built-in gravitational analog of a toroidal domain wall.
%%%%%%%%%%%%%%%%%%%%%%%%%%%%%%%%%%%%%%%%%%%%%%%%%%%%%
\section{Summary and discussion}
Here we presented a systematic derivation of Maxwell's equations in a stationary curved backgrounds using both threading and slicing spacetime decomposition formalisms, on the basis that they are formally equivalent to the Maxwell equations in flat spacetime in curvilinear coordinates. Using these equations we also obtained gravitationally-induced constitutive equations in both formalisms. The fact that these decomposition formalisms are assigned to two different sets of observers, the so called  fundamental, and fiducial observers, will help us to distinguish and compare the results of their measurements of different electromagnetic phenomena happening in stationary gravitational backgrounds. Of course if the effect is observer-independent one expects to find the same effect in both formalisms.
Specifically, the light ray trajectories are expected to be the same in both formalisms. On the other hand, light paths could be obtained through the Fermat's principle in stationary spacetimes which was shown to be equivalent in both formalisms. Indeed we arrived at the same definition for spacetime refractive index using different approaches, all based on the application of geometric optics in curved spacetime.\\
As an interesting application of the spacetime refractive index, it was shown that  its dependence on the direction of propagation will lead to the gravitational analog of non-reciprocal refraction in materials with toroidal domain wall. In this case the gravitational analog of the toroidal moment is nothing but the gravitomagnetic vector potential of the underlying spacetime. This was illustrated more clearly in the case of weak-field limit of stationary spacetimes, where it was shown that the gravitational analog of the relation between the toroidal moment, and spin in toroidal matter, is the similar relation between the gravitomagnetic potential and the angular momentum of the source particles. A very interesting example of this analogy is
shown to be provided by the NUT spacetime, which could be studied through ray-tracing simulation. Also to see the effects of both rotation and NUT parmeters at the same time, one can study the above-mentioned nonreciprocity in the Kerr-NUT solution where both parameters appear in the gravitomagnetic vector potential, and hence in the spacetime's index of refraction \cite{MEN}. Of course in this case it may not be expected to see a direct analogy with a toroidal domain wall, as was observed in the case of NUT solution.
%%%%%%%%%%%%%%%%%%%%%%%%%%%%%%%%%%%%%%%%%%%%%%%%%%%%5
\section *{Acknowledgments}
The authors would like to thank University of Tehran for supporting this project under the grants provided by the research council.
%%%%%%%%%%%%%%%%%%%%%%%%%%%%%%%%%%%%%%%%%%%%%%%%%%%%%%
\appendix
\section{Levi-Civita pseudotensors in 3D curved spaces}
In this appendix we give the relation between the Levi-Civita pseudotensors in flat 3D space, and the 3D curved spaces $\Sigma_3$, and $\Sigma_t$ in threading and slicing decomposition formalisms respectively.
Starting from the 4D general relation,
\begin{equation}
\eta_{\alpha\beta\mu\nu} = \sqrt{-g}~\epsilon_{\alpha\beta\mu\nu},
\end{equation}
in which $\epsilon_{\alpha\beta\mu\nu}$ is the anti-symmetric pseudotensor in flat spacetime with $\epsilon_{0123}=1$, the relation between 3D and 4D Levi-Civita forms in the two decomposition formalisms are given as follows: \\
{\bf 1})-In threading formalism this is done by using the 4-velocity of a fundamental observer such that
\begin {equation}\label{LCT}
\eta_{\alpha\beta\mu} = \eta_{\alpha\beta\mu\nu} u^\nu = \eta_{\alpha\beta\mu\nu}~\frac{\delta^\nu_0}{\sqrt{g_{00}}} = \frac{\sqrt{-g}}{\sqrt{g_{00}}}~\epsilon_{\alpha\beta\mu\nu}~\delta^\nu_0 = \sqrt{\gamma~}\epsilon_{\alpha\beta\mu 0}.
\end {equation}
where we made use of second relation in \eqref{gg}. Now the indices ${\alpha\beta\mu}$ could only take spatial values, so we end up with
\begin{equation}
\eta_{\alpha\beta\mu} \equiv \eta_{abm}=\sqrt{\gamma}~\epsilon_{abm}.
\end{equation}
Also, using the covariant 4-velocity of a fundamental observer, one can show that,
\begin{equation}
\eta^{abm}=\frac{1}{\sqrt{\gamma}}~\epsilon^{abm}.
\end{equation}
{\bf 2})-In the slicing formalism, similar result could be obtained for the 3D Levi-Civita tensor starting from the simpler case of the contravariant form of the above-mentioned general 4D relation, namely
\begin {equation}\label{LCS}
\eta^{\alpha\beta\mu} = \eta^{\alpha\beta\mu\nu}~u_\nu = \eta^{\alpha\beta\mu\nu}~N~\delta^0_\nu  = \frac{N} {\sqrt{-g}}\epsilon^{\alpha\beta\mu\nu}~\delta_\nu^0 = \frac{1}{\sqrt{\gamma}}\epsilon^{\alpha\beta\mu 0}.
\end {equation}
where we have used the 4-velocity of a fiducial observer \eqref{vel2}, and the second relation in \eqref{extras}. Again here the indices ${\alpha\beta\mu}$ could only take spatial values, and so
\begin{equation}
\eta^{\alpha\beta\mu} \equiv \eta^{abm}=\frac{1}{\sqrt{\gamma}}~\epsilon^{abm}.
\end{equation}
Using the contravariant 4-velocity of a fiducial observer, one can show that
\begin{equation}
\eta_{abm}=\sqrt{\gamma}~\epsilon_{abm}
\end{equation}
Indeed the relation between the 3D Levi-Civita pseudo tensor in flat and curved spacetimes is the same in both decomposition formalisms, though it is noted that the determinant of the 3D metrics ($\gamma $) included in the above definitions are defined in two different 3-spaces $\Sigma_3$, and $\Sigma_t$ for threading and slicing formalisms respectively. \\
Employing the above-defined Levi-Civita tensor in a curved 3D space, one can define  the curl and divergence of a vector {\bf A}, as well as its cross product with another vector {\bf B} as,
\begin{gather}
\nabla.\textbf{A}=\frac{1}{\sqrt{\gamma}}~\frac{\partial}{\partial{x^i}}(\sqrt{\gamma}~A^i) \label{div} \\
(\nabla \times \textbf{A})^i=\frac{1}{2}~\eta^{ijk}
(\partial_j{A_k}-\partial_k{A_j}) \label{curl} \\
({\bf A} \times {\bf B})^i = \eta^{ijk} A_j B_k \label{cross}
\end{gather}
%\pagebreak
%%%%%%%%%%%%%%%%%%%%%%%%%%%

\end{document}